# The *ECLAIRs* micro-satellite for multi-wavelength studies of gamma-ray burst prompt emission

S. Schanne, J.-L. Atteia, D. Barret, S. Basa, M. Boer, B. Cordier,
F. Daigne, A. Ealet, P. Goldoni, A. Klotz, O. Limousin, P. Mandrou,
R. Mochkovitch, S. Paltani, J. Paul, P. Petitjean, R. Pons, G. Skinner

*Abstract*--The cosmological revolution of 1997 has established that (at least long duration) gamma-ray bursts (GRB) are among the most energetic events in the Universe and occur at cosmological distances. The ECLAIRs micro-satellite, to be launched in 2009, will provide multi-wavelength observations for astrophysical studies of GRB and for their possible use as cosmological probes. It is expected to be the only space borne GRB trigger available for ground based robotic telescopes operational at that time.

This paper presents the ECLAIRs project and its status. An X/gamma-ray camera onboard ECLAIRs with a wide field of view of 2 sr, will detect ~100 GRB/yr in the 4-50 keV energy range, localize the GRB with a precision of ~10 arcmin on the sky, and transmit this information to the ground in near real-time, as a GRB trigger for ground based optical telescopes. Inspired by the INTEGRAL imager IBIS, it is based on a CdTe detection plane covering 1000 cm$^2$, placed 35 cm below a coded mask. An optical camera, sensitive to magnitude-15 stars, covering up to 1/4th of the X/gamma-ray camera's field of view, will observe the prompt emission and a possible precursor of ~10 GRB/yr in the visible-band. Used in a continuous acquisition mode at a rate of ~5 images/s dumped into an on-board memory, a GRB event sent by the X/gamma-ray camera triggers a seek-back in memory for the GRB optical precursor. The full X/gamma-ray and visible-band data of a GRB are sent to ground when a high data-rate telemetry ground receiver is reachable.

## I. INTRODUCTION

Gamma-ray bursts (GRB), revealed in space borne gamma-ray detectors by an important count rate increase for a short duration, remained very puzzling high-energy phenomena for a long time after the revelation of their discovery in the 1970's. The BATSE detector onboard the NASA CGRO satellite provided a first firm hint on the cosmological origin of GRB, based on a 9 year-map showing 2704 GRB distributed uniformly on the sky [1]. Statistical studies show furthermore that about 80% of all GRB last less than one minute, and that there is a class of short-duration GRB which last less than a few seconds. The real breakthrough in understanding GRB came with the discovery of X-ray [2], optical [3], and radio [4] GRB afterglows. The detection of the host galaxies and the measurement of their redshift showed that (at least long-duration) GRB are events related to the final stage in the evolution of massive stars, taking place at cosmological distances. The gamma-rays of a GRB are believed to be produced by internal shocks taking place in a collimated jet directed close to the line of sight, while the afterglow would result from the interaction of the jet with the surrounding medium. The total gamma-ray energy radiated by a GRB amounts to more than $10^{51}$ erg (while the energy released in the visible-band for a typical supernova is about $10^{49}$ erg), making GRB the most energetic explosions known in the Universe since the Big Bang.

With operations starting in 2009, the space-borne GRB detector ECLAIRs will offer multi-wavelength observations of about a hundred GRB per year, covering simultaneously the visible and the X/γ-ray domain, and will contribute to the rich field of research which emerged from GRB observations.

## II. SCIENTIFIC OBJECTIVES FOR GRB STUDIES

### A. *GRB observations for cosmology*

The study of gamma-ray bursts offers important perspectives for the progress of modern cosmology. Indeed, due to their high intrinsic luminosity, GRB are well suited for probing the Universe to very high redshifts (z up to 10-15, if GRB exist at those high z). As in the case of quasars, the interstellar medium of GRB host-galaxies and the intergalactic medium are probed by the GRB. Their properties are imprinted in the GRB afterglow. In contrary to quasars, GRB can be well studied, since the GRB host galaxies are neither very active, nor very massive, and their environment is not ionized by the radiation of the GRB. The hypothesis that a GRB is produced during the gravitational collapse of a very massive star is confirmed by the fact that some GRB (if not all of them) are associated with a particular class of supernovae of type Ic (SN Ic), sometimes called hypernovae [5], [6]. The GRB rate as a function of z becomes therefore a direct tracer of the star formation rate, in particular at high z. It becomes even possible, using GRB, to detect the first generation of stars (population III), which could have been particularly massive, and responsible for the re-ionization of the Universe and an early synthesis of metal elements. In this context, ECLAIRs will provide in near real-time the position of GRB to large ground-based telescopes. Such an early trigger is crucial to their ability to perform detailed spectral analyses of early GRB afterglows.

*B. Studies of GRB physics*

The study of the astrophysical phenomenon of GRB will be pushed forward by ECLAIRs. Indeed, GRB are not only extreme events by the amount of released energy, but also by the physical processes at work. According to the most studied model for long duration GRB, the collapsar model (Fig. 1), a GRB is a consequence of the core collapse of a very massive star, with the formation of a black hole in its center followed by relativistic ejection of matter. In this model, the extremely high γ-ray flux is produced by particles accelerated in successive shocks which occur inside the relativistic wind itself. However, much less is known concerning the origin of the prompt emission in the X-ray and visible band. The prompt visible emission, detected only once (in the case of GRB 990123 [7]), could be the consequence of a reverse shock inside the relativistic wind generated during its interaction with the interstellar medium. The prompt X-ray emission could be the result of internal or external shocks, as it is the case for the γ-ray emission. However it could also be thermal emission released when the relativistic ejecta becomes transparent. In some models the X-ray emission could be emitted prior to the γ-rays. In order to understand the physics of GRB, in particular their central engine, observations in a spectral domain as wide as possible and over a time-span as long as possible − ranging from a possible precursor to the transition between prompt and afterglow emission − are necessary. In this respect, thanks to its capability of observing in the γ-ray, X-ray and visible domain right from the beginning of the event, ECLAIRs will provide a better understanding of the physical processes at work in GRB. Those processes involve among others complex hydrodynamic processes (relativistic outflows and shocks, also at work in AGN and microquasars) and emission processes (thermal, synchrotron, inverse Compton, pair annihilation, etc). ECLAIRs will also help to determine the nature of GRB progenitor stars. No other scientific instrument, in operation or planned, will be able to observe in the visible and X-ray band the prompt emission of GRB during the first 60 s after the event, which is precisely the period of time in which GRB produce their γ-rays (in 80% of the cases). Studies of the physical processes involved in GRB are not only important in order to understand the GRB phenomenon itself, but also to determine, if it could ever be possible to consider some of those events as "standard candles" for cosmological surveys.

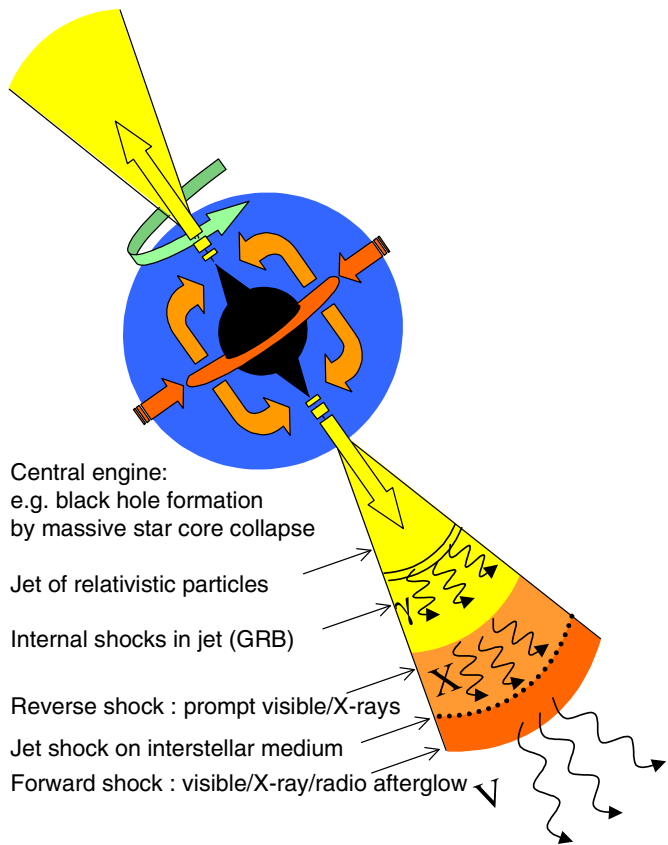

Fig. 1: A schematic model for long-duration GRB: the collapse of a rotating massive star (collapsar). When the nuclear fuel in the star core is exhausted, the core collapses. A black hole forms, surrounded by a thick plasma disk (in the equatorial regions of the star) and an outward flow of relativistic particles (in the polar regions). In-falling matter may excavate a channel in the polar regions, from which relativistic particles can be ejected with Lorentz factors > 100 in the form of a jet with an opening angle of a few degrees. Inside the jet, shocks between fast moving shells of particles encountering slower moving shells, result in particle acceleration and emission of gamma-rays (gamma-ray burst). At contact of the jet with the interstellar medium, a shock propagates into the interstellar medium and produces the afterglow observable in the visible, X-ray and radio bands. A reverse shock inside the jet may be responsible for the emission of prompt X-rays and visible light.

### III. FUTURE STRATEGIES FOR GRB STUDIES

The French astrophysical community, with others, has been actively involved for several years in a three step strategy for the observation and study GRB (Fig. 2). Those steps are: (i) the detection of the GRB using space-borne gamma-ray detectors capable of determining in real-time the localization of the event in the sky with an accuracy of a few arc-minutes; (ii) the prompt observation with small robotic telescopes of the error boxes provided by space instruments, with a view to detecting the GRB afterglow in the visible and near-infrared and determine the localization of the event with arc-second accuracy; (iii) spectroscopic observation of the afterglow emission, as quickly as possible, with large ground based telescopes in the visible and space borne telescopes in X-rays.

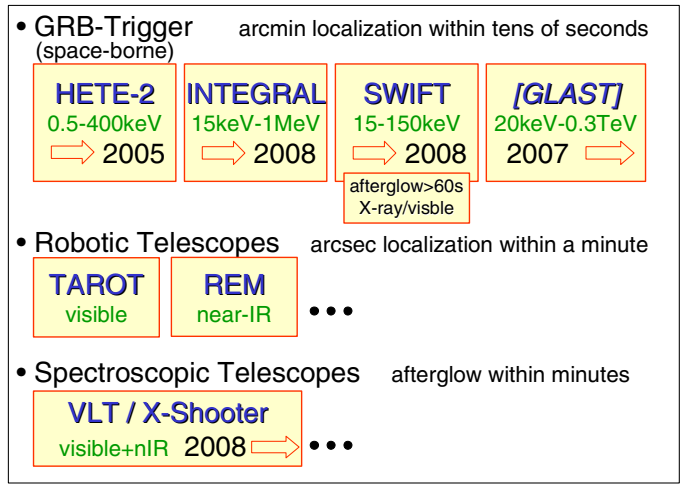

Fig. 2: Three step strategy for GRB observations. The GRB is detected with a space-borne trigger and localized within an arcmin error-box in near-realtime; gound-based robotic telescopes localize the event with arcsec accuracy; ground-based spectroscopic follow-up observations of the visible afterglow within minutes after the event are planned. However, this picture shows two deficiencies: (i) after 2008 no space-borne GRB-trigger is available any longer (GLAST will not reach the arcmin localization precision for most GRB); (ii) the GRB prompt emission in the visible and X-ray band is not studied with the presently available or planned instruments. The aim of ECLAIRs is to contribute to both of those topics.

For the success of this program, many scientists are currently involved in the first step with the scientific data analysis of the HETE-2 [8] and INTEGRAL [9] satellites, which are capable of localizing in near real-time GRB with a sufficient accuracy in order to allow ground-based follow-up observations. Many scientists are also involved in the second step, for which the duplication of the TAROT robotic telescope [10] in Chile is ongoing, and the commissioning of the robotic telescope REM for observations in the visible and near infrared [11] is currently progressing at la Silla in Chile. Many other robotic telescopes are in use or under construction all around the world, in particular in Europe, the US and Asia. For the third step, the European scientific community contributes to the X-SHOOTER project [12], a second generation instrument at the VLT of the European Southern Observatory in Chile, aiming at spectroscopic studies of GRB afterglows in the visible and near infrared band, with observations starting between 2007 and 2008. This instrument will permit to study the environment of the GRB progenitor, the interstellar medium of its host galaxy and the intergalactic medium in the line of sight. For these studies, it is important to get as input frequent and fast positions of gamma-ray bursts. In the present planning, and in the absence of ECLAIRs, there will be no space-borne GRB-detector capable of providing this information. The ECLAIRs mission, which is designed to detect GRB at a relatively high rate and localize them with a good accuracy, appears to be the only mission able to provide GRB alerts for X-SHOOTER, under the condition that its launch takes place shortly after the X-SHOOTER first light.

In the international context, the NASA gamma-ray burst mission SWIFT [13], with a launch date scheduled for end of 2004, has been built to detect more than 100 GRB per year and to observe the afterglow emission in visible and X-rays about 60 s after the gamma-ray event. This time is needed in order to point the narrow field instruments of SWIFT (X-ray and optical cameras) to the GRB position determined by the large field γ-ray telescope onboard SWIFT. Therefore SWIFT is often unable to observe the prompt emission of the GRB in the visible band and in X-rays. Furthermore the SWIFT γ-ray detector, foreseen to operate between 15 and 150 keV, has a restricted energy-band and will be able to characterize precisely the spectral properties (among which $E_{peak}$) of only a small fraction of GRB.

The only other gamma-ray astronomy mission planned for operation after the year 2007 is the NASA high-energy gamma-ray astronomy satellite GLAST [14]. For GRB alerts, GLAST is equipped with the gamma-ray burst monitor GBM [15], which has an energy-band from 10 keV to 20 MeV. Based on the design of BATSE, the GBM will be able to detect about 100 GRB per year, however it will not be able to produce localization better than a few degrees, preventing ground based follow-up observations of the afterglow. Furthermore the GLAST main instrument LAT has an energy band starting at 20 MeV, ranging up to hundreds of GeV, and will therefore be able to localize precisely only those GRB which emit a sufficient photon flux at high energies, limiting and biasing the population of GRB for which the LAT delivers a trigger.

The launch of ECLAIRs is foreseen shortly after the commissioning of GLAST and both instruments have large fields of view. Thus a fraction of the GRB seen at high-energies by the LAT of GLAST will also be observed by ECLAIRs. Those GRB observed simultaneously by GLAST and ECLAIRs will be studied over a very large spectral domain, from visible to high-energy γ-rays. The complementary spectral coverage offered by ECLAIRs will help in understanding the high-energy emission of the GRB observed by GLAST. Furthermore, ECLAIRs will provide the precise localization of the GRB detected by GLAST, and therefore permit ground based follow-up observations of GRB emitting at high-energies.

IV. SCIENTIFIC SPECIFICATIONS OF THE ECLAIRs MISSION

The main scientific specifications of the ECLAIRs mission are the following. (i) Detect about 100 GRB per year, independently of their duration, detect in particular the short duration GRB, the X-ray rich GRB and the X-ray flashes; (ii) compute in near real-time the position of the GRB on the sky with an accuracy of up to 10 arcmin, transmit this information on-ground in real-time and distribute it as fast as possible to other observatories (ground-based or space-borne);

(iii) observe the prompt emission and any precursor to the GRB (using on-board storage capabilities) in the γ-ray, X-ray and visible bands and transmit this information on-ground after some delay.

The GRB imaging device of ECLAIRs is inspired from the very successful coded mask aperture telescope IBIS on-board INTEGRAL, ESA's γ-ray astrophysical laboratory, in orbit since October 2002. The IBIS/ISGRI imaging detection plane, built at CEA Saclay, France, is composed of an array of 16384 CdTe semiconductor crystals. This detector is very efficient in detecting and localizing GRB: despite a moderate field of view (9°×9° fully coded), IBIS detects about 1 GRB per month with a localization accuracy of about 1 arcmin. The INTEGRAL satellite has a permanent data-link with the ground, which permits the downlink of data from all INTEGRAL detectors with minimal delay. The scientific data are collected at the INTEGRAL Science Data Center (ISDC), Versoix, Switzerland, where a ground-based software-system (the INTEGRAL Burst Alert System IBAS) [16] is running in order to detect and localize GRB in near real-time, using the data of IBIS/ISGRI. In many cases, the system generates a GRB alert within tens of seconds after the GRB event, together with event localization with accuracy better than 2 arcmin.

In the framework of the ECLAIRs mission it is foreseen to utilize a device similar to IBIS/ISGRI together with a software system similar to IBAS on a micro-satellite platform. This is a challenge, which is nevertheless possible (1) by reducing the mask-detector distance in order to open the field of view to cover 2 sr of the sky, without degrading much the source localization capabilities (10 arcmin accuracy); (2) by choosing a low Earth orbit with the lowest possible inclination angle in order to reduce the particle induced γ-ray background of the mission; and (3) by using the 4-50 keV energy band for the GRB detection and localization, which, compared to the 15-200 keV band of IBIS, needs less thickness for the mask and the lateral shielding of the camera, and which covers the scientifically interesting X-ray rich GRB. Without loosing in sensitivity compared to INTEGRAL, ECLAIRs therefore has a smaller detection area, and its low mass (50 kg) and power consumption (45 W) makes it possible to fit the mission into the framework of a micro-satellite of the Myriade family, developed by the French Space Agency CNES.

## V. ECLAIRs MISSION EXPERIMENTAL SET-UP

The ECLAIRs experiment is designed as a payload (on a micro-satellite platform of the Myriade type) which is composed of 3 distinct subsystems mounted on one face of the (cubic-shaped) Myriade platform (Fig. 3): the Payload Interface Module (MCU), the Camera for X and Gamma-ray detection (CXG) and the Unit for Detection in the Visible-band (UDV) whose task is to observe up to a quarter of the CXG field of view in order to follow in the visible band the prompt emission and possibly the visible precursor of more than 10 GRB per year.

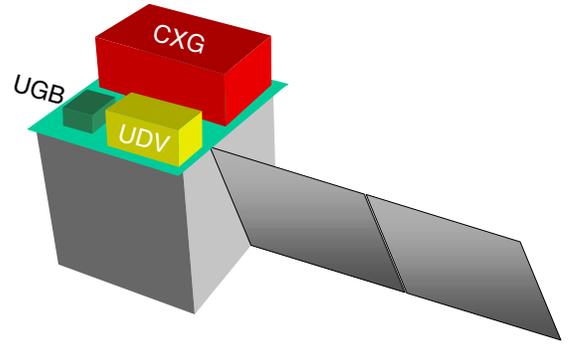

Fig. 3: The ECLAIRs payload on a Myriade micro-satellite platform.

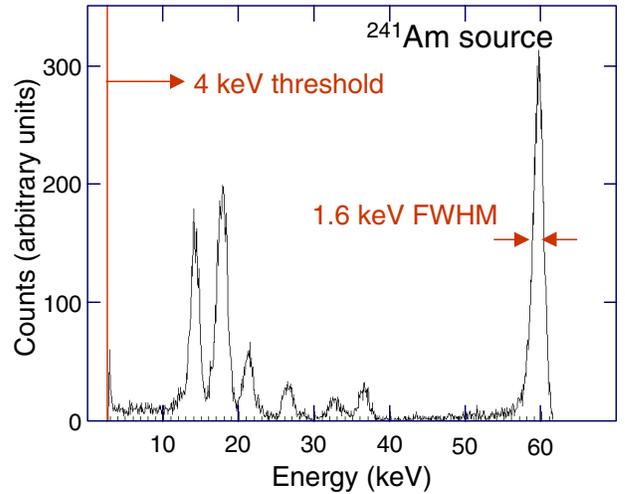

Fig. 4: Spectrum obtained in the lab [17] with an $^{241}$Am source illuminating a CdTe pixel produced by ACRORAD (Japan) and read-out by the IDeF-X chip developed at CEA-Saclay. The CdTe pixel (dimensions 2×2×0.5 mm$^3$, surrounded by a guard-ring) is reverse biased under 200 V, and its anode is equipped with a Schottky contact with a very low dark-current < 50 pA at room temperature. This device shows a very good spectral resolution: 1.6 keV (FWHM) is measured for the 59.5 keV line of source. The low energy threshold is below 4 keV, which demonstrates the feasibility of an integrated X-ray/γ-ray camera for ECLAIRs.

### A. The ECLAIRs X/γ-ray detector (CXG)

The CXG, to be developed by CESR, Toulouse, France, is a wide field coded-mask aperture telescope, with a detection plane which could be made of about 6000 CdTe crystals, with a size of 4×4 mm$^2$ and a thickness of 2 mm, surrounded by a guard ring of 0.5 mm and Schottky contacts. CdTe pixels are able to detect X-ray and γ-ray photons with high efficiency and good spectral performance. The leakage current is mainly evacuated into the guard ring, therefore leakage is very small inside the sensitive area. In order to reduce the noise, the contacts are connected to a preamplifier through a substrate with a low parasitic capacitance. Furthermore, due to the low leakage current, a high external electrical field (of about 400 V/mm) can be applied, the charge collection is almost complete and an excellent energy resolution is obtained. The industrial production technology of CdTe is nowadays very stable. In order to meet the low energy threshold (of 4 keV) and the low power consumption (below 1 mW/channel), the detectors are coupled to a chip called IDeF-X, developed at CEA Saclay, France, in the framework of a CEA-CNES R&D program. First tests show that the requirements can be met (results presented in [17] and Fig. 4). The CdTe detectors show an excellent efficiency at low energies (below 50 keV).

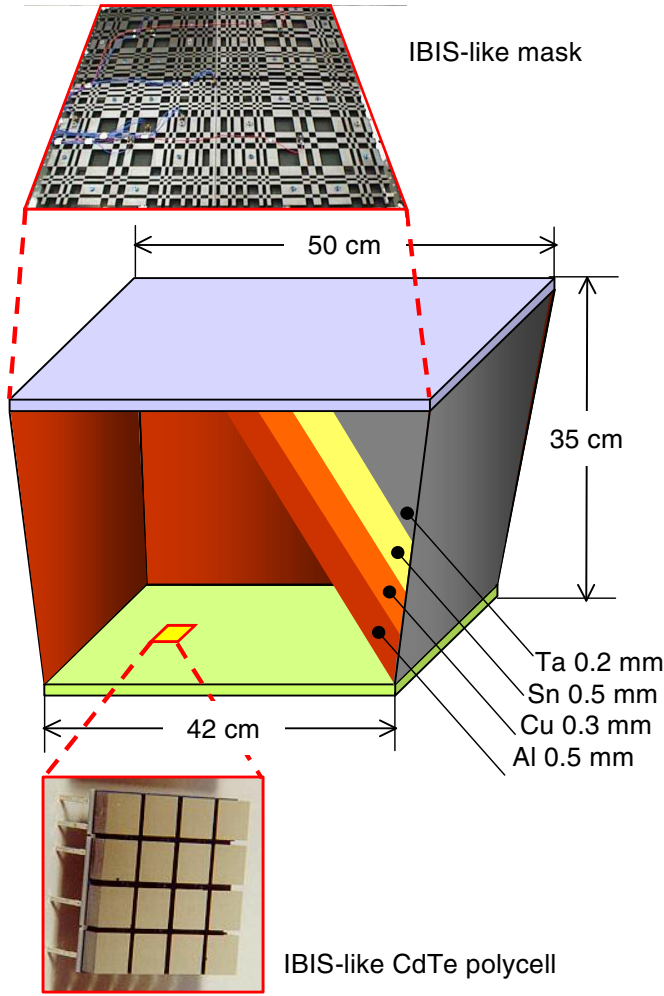

Fig. 5: Schematic view of the ECLAIRs CXG. A coded mask is placed 35 cm above the CdTe detection plane inspired from INTEGRAL/IBIS, surrounded by a passive shielding made of 4 layers of gamma-ray absorbing materials.

The current design of the CXG (Fig. 5) has a total field of view of about 2 sr (105°×105°) and is based on a detection plane with a sensitive area of about 1000 cm², placed 35 cm below a coded mask (50% transparency). The CXG is surrounded by a passive shield (opaque up to 50 keV, 20% transparent at 100 keV) made of successive Ta, Sn, Cu and Al foils (thickness 0.2-0.5 mm, foils listed inwards), one foil shielding the X-ray fluorescence of the previous one. A Monte-Carlo simulation by O. Godet (Fig. 6) shows that the localization accuracy for such a camera is 12 arcmin; a sensitivity of ~170 mCrab in the 4-50 keV band (with a signal to noise ratio of 5.5) is reached for an accumulation time of 10 s. Allowing for a 70% observing efficiency, based on the log N-log S distribution of GRB detected by BATSE and a mean emission spectrum of the GRB (broken power-low, Band model [18], with $\alpha=-1$; $\beta=-2$ and $E_{peak}=150$ keV), and a 40% fraction of X-ray rich GRB, we expect to localize with the CXG about 100 events per year in the 4 to 50 keV band.

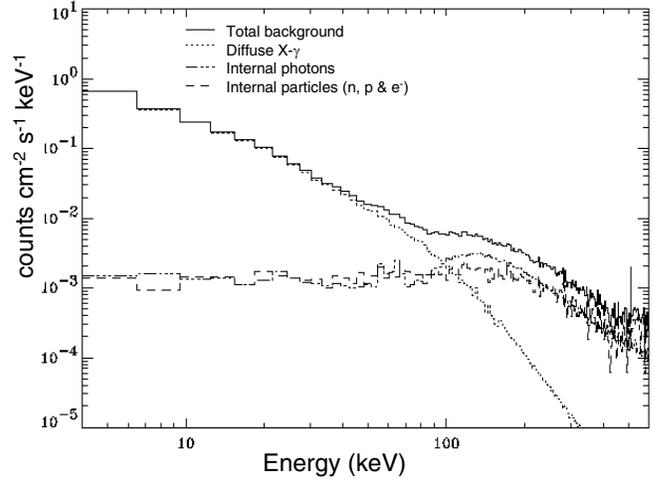

Fig. 6: Monte-Carlo simulation (GEANT) by O. Godet of the background expected for the ECLAIRs-CXG, taking into account the diffuse extra-galactic X/γ-rays, primary cosmic protons and electrons, trapping of particles in the Earth magnetic field, secondary protons, neutrons and photons from the Earth albedo. The orbit was chosen with an altitude of 600 km and an inclination of 20°.

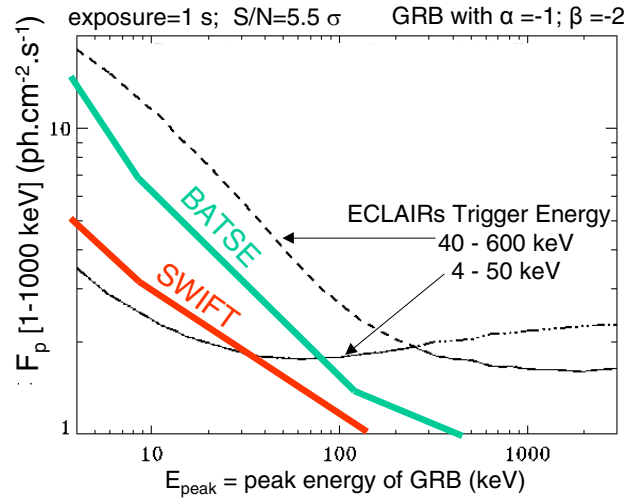

Fig. 7: GRB detection sensitivity (at a 5.5σ level, by O. Godet) for an exposure of 1 s as a function of the GRB peak-energy ($E_{peak}$). The GRB spectrum used follows a power low with index $\alpha=-1$ for $E< E_{peak}$ and $\beta=-2$ for $E>E_{peak}$. For each $E_{peak}$, the GRB spectrum is integrated over two ECLAIRs triggering domains (the nominal 4-50 keV range, and a 40-600 keV range) and compared to the expected background in the same band; the detection sensitivity is derived (GRB integral flux in the 1-1000 keV range). With a low energy triggering threshold of 4 keV, we expect for ECLAIRs a better sensitivity than BATSE for GRB with $E_{peak}<70$ keV, and than SWIFT for GRB with $E_{peak}<40$ keV.

## B. The ECLAIRs detector for the Visible-band (UDV)

The UDV, to be developed jointly by LAM, Marseille, France, and OHP, Saint Michel l'Observatoire, France, is an optical system covering a field of view of up to 40°×40°, aligned with the optical axis of the CXG, made of up to 4 independent cameras. Each camera is equipped with a CCD. One type of CDD under consideration is EEV4280 with 2048×4096 pixels, also used for the COROT satellite. This low current, passively cooled CCD covers the 370-950 nm band (suitable for red shifted objects, Fig. 8). No filter and no cryogenic cooling are foreseen. Half of the CCD area is hidden, and used as a buffer in order to avoid the usage of a mechanical shutter, reducing the dead-time of the camera. An image is transferred in about 200 ms into the hidden-layer from where it is read-out in less than 4.4 s. Therfore the minimal exposure time is less than 5 s. In the case of a GRB trigger at a given position inside the UDV field, a sub-window of the CCD can be read-out at a higher rate, which allows a fine-time follow-up of the GRB visible counterpart. The UDV electronics is similar to the one developed at LAM for the COROT mission. The intrinsic camera resolution is 30 arcsec, but the resolution is limited by the satellite pointing stability (better than 5 arcmin). The diameter of the lens of each camera is about 30 mm. A 10σ-sensitivity for detecting an object of magnitude 15 (in V-band equivalent) is reached with an exposure of less than 30 s, in case of no confusion with a neighboring bright star (Fig. 9). Apart from GRB observations the UDV cameras are also used for the satellite attitude control. Due to its wide field-of-view, a scientific byproduct of the UDV will be the routine sky-surveillance in the visible for objects (dis-)appearing on short time-scales.

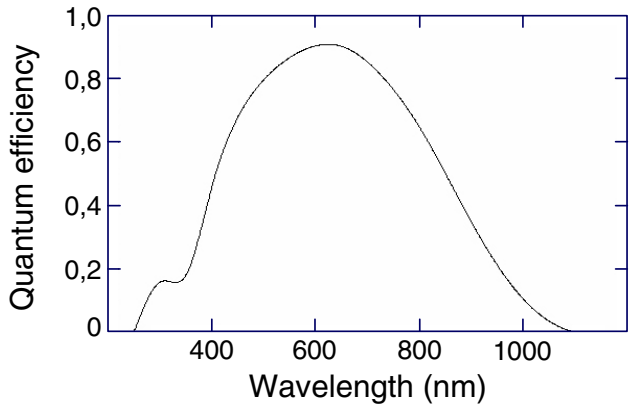

Fig. 8: The quantum efficiency as a function of the visible-light wavelength for the CCD EEV 4280 (also selected for COROT) shows that spectral coverage into the red-light domain is achieved, favorable for detection of red-shifted objects at cosmological distances.

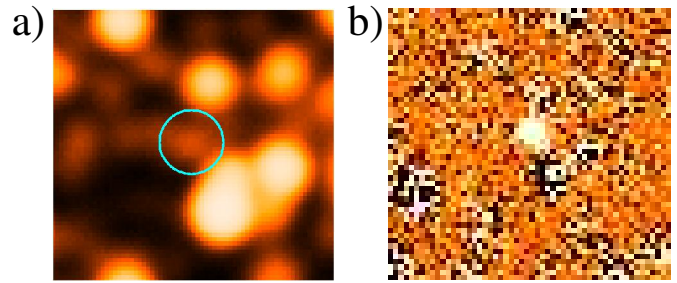

Fig. 9: Simulation of the observation of a V-band magnitude-15 star in a 30 s exposure in case of no source confusion (a). The source is detected in (b) after subtraction of the exposure from the same sky-region in which the source is absent. Size shown: 52×52 pixels (30×30 arcmin$^2$).

## C. The ECLAIRs payload module and electronics (MCU)

The Payload Interface Module (MCU, *Module Charge Utile*), to be developed by DAPNIA, CEA-Saclay, France, is dimensioned to fit onto one face of the micro-satellite. The CXG and UDV detectors are mechanically integrated onto the MCU, as well as the payload Board Electronics Unit (UGB, *Unité de Gestion de Bord*). The UGB (Fig. 10) is in charge of the electrical power supply to the cameras, their command/control and their data acquisition, storage and processing, among which the most important aspect is the GRB detection and localization.

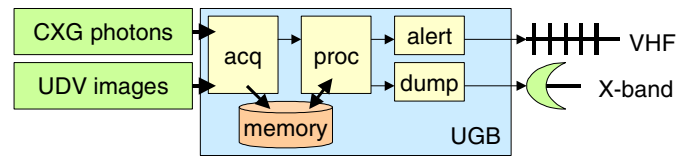

Fig. 10: ECLAIRs data acquisition system. The UGB acquires continuously photon packets from the CXG and visible images from the UDV, stores them into memory and computes simultaneously the GRB trigger conditions. A count rate-increase is measured at the data acquisition stage (acq, which could be implemented in FPGA), while a gamma-ray source position is computed in a software based processing unit (proc). The UGB manages the ground communication via VHF and X-band.

The UGB receives from the CXG a continuous stream of photon packets (per photon: energy, location on the camera, time) and from the UDV a sequence of images (about 1 every 5 s). A time stamp is attached to each data packet.

With its computing capabilities the UGB routinely determines the spacecraft pointing direction by comparing the UDV sky images to an onboard star catalogue. Differences of UDV sky image are routinely built to search for variable sky objects in the visible. The parameters (date, position, magnitude) of excesses are stored in an internal memory table.

The UGB performs the trigger on GRB events by computing in real-time the CXG counting rate in several energy bands (between 4 and 50 keV). A significant count-rate increase issues a GRB trigger of a first type. In parallel, using the CXG photons contained in a sliding time-window (duration a few s), the UGB continuously tries to find any previously unknown γ-ray source on the sky (by deconvolution of the mask-pattern projected by the source on the CXG detection plane), in which case a GRB trigger of a second type is issued. The second kind of trigger may reveal a GRB too weak to be detected by the count rate increase trigger, in particular X-ray rich GRB and GRB at high redshift. In both triggering cases, the UGB sends a quick alert to one of the ground stations of the permanently visible network (Fig. 11) via a low bit-rate VHF telemetry stream. In case of a count rate increase trigger, the image deconvolution process is also run in order to localize the event. The ground alert contains the type of the event, the start time of the event, confidence level, and (whenever possible) position and error box. The ground receiver-system subsequently transmits this information as quickly as possible to the observing community.

Once a GRB candidate is localized, the UGB defines a window for the UDV camera containing the region of interest, which is subsequently read-out at a higher rate (typically every 1 s). The GRB candidate is followed-up for several minutes, and the CXG and UDV data are stored. The memory containing CXG and UDV data for several minutes before the GRB candidate is also stored, since they may contain GRB precursor information.

The UGB mass memory data are transmitted to the ground whenever the satellite is in contact with a ground station supporting the X-band high bit-rate data transfer, which occurs at least once per day. It is foreseen to store onboard all the CXG photons acquired, in order to perform X-ray/$\gamma$-ray source studies as a scientific byproduct in the absence of GRB. In case of memory shortage, the GRB data have storage priority.

The UGB electronics uses FPGA for the CXG and UDV dialogue and microprocessors for the image deconvolution. Both the FPGA configuration and the processor code can be replaced by new versions sent from the ground.

## VI. ECLAIRs MISSION PROFILE

The ECLAIRs mission aims at a launch date in 2009. In order to limit the costs, the satellite could be launched as a passenger of another satellite whose launch date fits in our planned time-frame. A low Earth orbit (altitude between 500 and 900 km, ~90 min orbital period) is foreseen for the Myriade micro-satellite family, well suited for ECLAIRs in order to operate below the Earth radiation belts, avoiding a high radiation dose. A low inclination orbit is desirable, in order to avoid the South Atlantic Anomaly in the Earth magnetic field, where the radiation level is increased, and to facilitate continuous ground contact.

### A. ECLAIRs pointing strategy

The platform is foreseen to be operated in a 3 axis-stabilized pointing mode to avoid star motion across the UDV cameras during observations. The pointing direction is programmed in advance to avoid the Sun, the Earth, the Moon and bright planets in the UDV field of view; the fraction of the Earth in the CXG field of view is minimized. Furthermore the pointing is optimized to keep always the same face of the satellite oriented towards the cold (anti-solar) for thermal stability. The galactic plane is avoided if possible in order to limit the number of stars in the UDV view. Pointing directions for which the filed of view would be accessible to visible or infra-red telescopes in the night-zone on the ground will be preferred.

### B. ECLAIRs ground-segment

The ECLAIRs ground-segment (Fig. 11) is inspired from the one developed for the Myriade micro-satellite DEMETER (launched in June 2004) with a Scientific Mission Center (CMS) closely linked to the Mission Control Center (CCM) at CNES. The CCM is in charge of sending commands to the satellite and receiving the raw data form the CNES receivers. The CMS is in charge of monitoring the performance of the payload, programming the payload in-flight operations and managing the scientific data. This includes the data pre-processing, scientific analysis, archiving and distribution to the involved scientific institutes, as well as the distribution of the GRB alert to the observing community in near real-time (within tens of seconds) via internet and e-mail.

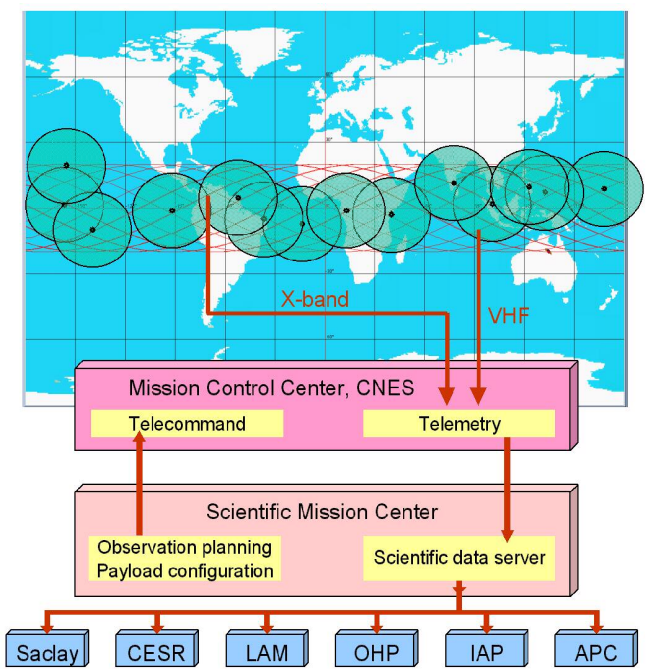

Fig. 11: A network of VHF ground stations (inspired from the HETE-2 one) is used to receive efficiently the near-real time GRB alerts from ECLAIRs (orbit drawn with e.g. inclination 20°, altitude 866 km). Scientific data are transmitted to an X-band receiver (>16 Mbit/s) when the ground station is visible. Satellite control and telemetry reception are managed by the Mission Control Center. At the Scientific Mission Center the ECLAIRs observation planning and configuration are defined, the scientific telemetry is processed, and the data server for the scientific community is maintained.

### C. ECLAIRs telemetry streams

The positions of the GRB detected by ECLAIRs are sent to ground via a VHF radio transmitter to a network of receivers located along the equator. In this respect the orbital parameters will be chosen such as to allow a permanent VHF contact with at least one ground receiver for 90% of the time. The ground-receiver network could be similar to that developed for HETE-2 with an extension to higher latitudes depending on the final orbit configuration.

The continuous VHF telemetry stream from ECLAIRs is composed of technological housekeeping data in absence of GRB. If a GRB occurs, the content of the message is changed and incorporates the most up-to-date information available about the GRB (type, time, confidence level, localization on the sky, error box). This message, refined if possible, is continuously repeated, so that it can be received by more than one VHF ground station for redundancy purposes and to manage hand-over between VHF stations.

The complete set of scientific data is transmitted via a high bit-rate X-band telemetry stream whenever the X-band ground station (possibly located at Kourou, French Guyana) is visible. It contains the CXG and UDV data acquired minutes before and after the GRB. In absence of GRB it contains (whenever possible) all CXG photons since the last memory dump via X-band, a subset of optical-camera images and the list of variable objects in the visible band.

## VII. THE ECLAIRs MISSION SCHEDULE

The ECLAIRs mission is built using a reduced cost development scheme, which avoids the successive intermediate models before the flight-model construction and uses cost effective radiation tolerant components. The mission is built for a lifetime of 2 years.

The ECLAIRs concept previously presented [19] has been altered and further developed. Project management is now at DAPNIA, CEA Saclay, France. Preliminary mission design studies have been completed. The mission has passed a critical step in the process of selection by CNES. A phase-A study will start in 2005 with a detailed design of the mission subsystems. The development of the ECLAIRs subsystems is planned between 2006-2007, the mission integration and ground-based tests are foreseen in 2008. The launch is scheduled in 2009. This launch date corresponds to the end of operations for INTEGRAL and SWIFT. Therefore ECLAIRs will be the only instrument capable to provide GRB triggers after 2009, right at the moment when large ground based robotic telescopes equipped with powerful instruments (like X-SHOOTER at the VLT) will be operational.

## VIII. ACKNOWLEDGMENT

The authors would like to thank O. Godet from CESR, Toulouse, France, for providing results and figures prior to his Ph. D. defense. The authors would also like to thank M.-A. Clair and her colleagues from the CNES Micro-Satellite Division in Toulouse, France, as well as the DEMETER team from Orléans, France, for very fruitful discussions.

## IX. REFERENCES

[1] W. S. Paciesas 2004, Baltic Astronomy, vol. 13 pp. 187-192 & W. S. Paciesas, C. A. Meegan, G. N. Pendleton et al 1999, ApJS 122, 465
[2] E. Costa, F. Frontera, J. Heise et al 1997, Nature 387, 783
[3] J. van Paradijs, P.J. Groot, T. Galama et al 2003, Nature 386, 686
[4] D.A. Frail, S.L. Kulkarni, S.R. Nicastro et al 1997, Nature 389, 261
[5] J. Hjorth, J. Sollerman, P. Meller et al 2003, Nature 423, 847
[6] K. Z. Stanek, T. Matheson, P. M. Garnavich et al 2003, ApJ 591, L17
[7] R. Sari & T. Piran 1999, ApJ 517, L109
[8] G. R. Ricker, J.-L. Atteia, G. B. Crew et al, Gamma-ray Bursts and Afterglows, ed. Ricker & Vanderspek, 2001, *AIP Conf. Proc.* 662, 3
[9] P. Ubertini, F. Lebrun, G. Di Coco et al 2003, A&A 411, L131
[10] M. Boer & A. Klotz 2003, *GCN Circ.* N° 2188
[11] F. Vitali, F. M. Zerbi, G. Chincarini et al 2003, *Proc. of SPIE*, vol. 4841 pp. 627-638
[12] A. Moorwood & S. D'Odorico, March 2004, ESO Messenger 115, p. 8, & http://www.eso.org/instruments/xshooter/
[13] N. Gherels, 2001, in "Gamma-Ray Bursts in the Afterglow Era", ed. E. Costa, F. Frontera & J. Hjorth (Berlin, Heidelberg, Springer), 357
[14] G. G. Lichti, M. Briggs, R. Diehl, et al 2004, astro-ph0407137
[15] A. von Kienlin, C. A. Meegan, G. G. Lichti, et al 2004, astro-ph0407144 & G. G. Lichti, M. S. Briggs, R. Diehl et al 2003, *Proc. of SPIE*, vol. 4851 pp. 1180-1187
[16] S. Mereghetti, D. Götz, J. Borowski et al 2003, A&A 411, L291
[17] O. Limousin, O. Gevin, F. Lugiez, B. Dirks, R. Chipaux, E. Delagnes & B. Horeau, *Proc. of IEEE NSS conf.*, Rome, 2004 (to be published)
[18] D. L. Band, Burst Populations and Detector Sensitivity, to appear in *Proc of the 2003 GRB conference* (Santa Fe, 2003, Sep 8-12), astro-ph/0312334.
[19] B. Cordier, J. Paul, D. Barret, G. Skinner, J.-L. Atteia & G. Ricker 2003, *Proc. of SPIE*, vol. 4851 pp. 1188-1195